\documentclass[12pt,a4paper]{article}
\usepackage[utf8]{inputenc} % allow utf-8 input
\usepackage[T1]{fontenc}    % use 8-bit T1 fonts
\usepackage{hyperref}       % hyperlinks
\usepackage{url}            % simple URL typesetting
\usepackage{booktabs}       % professional-quality tables
\usepackage{amsfonts}       % blackboard math symbols
\usepackage{nicefrac}       % compact symbols for 1/2, etc.
\usepackage{microtype}      % microtypography
\usepackage{xcolor}         % colors
\usepackage{microtype}
\usepackage{graphicx}
\usepackage{subfigure}
\usepackage{bbm}
\usepackage{booktabs} % for professional tables
\usepackage{algorithm2e}
\usepackage{algorithm}
\usepackage{multicol}
\usepackage{natbib}
\usepackage{algpseudocode}
\algrenewcommand\algorithmicrequire{\textbf{Input:}}
\algrenewcommand\algorithmicensure{\textbf{Output:}}
\usepackage{hyperref}

\usepackage{amsmath}
\usepackage{amssymb}
\usepackage{mathtools}
\usepackage{amsthm}
\usepackage{bm}
\usepackage{natbib}
\theoremstyle{plain}
\newtheorem{theorem}{Theorem}[section]
\newtheorem{proposition}[theorem]{Proposition}

\theoremstyle{definition}
\newtheorem{definition}[theorem]{Definition}

\theoremstyle{remark}
\newtheorem{remark}[theorem]{Remark}
\usepackage{pifont}
\newcommand{\cmark}{\ding{51}}%
\newcommand{\xmark}{\ding{55}}%
\newcommand{\equalcontribmark}{%
{\renewcommand\thefootnote{\fnsymbol{footnote}}\footnotemark[1]}%
}
\newcommand{\equalcontribtext}{%
{\renewcommand\thefootnote{\fnsymbol{footnote}}\footnotetext[1]{ Contribute equally to this work.}}%
}
                 
\def\Rb{ {\mathbb R}}
\def\E{ {\mathbb E}}

\def\I{ {\mathcal I}}

\def\P{ {\mathbb P}}   

\def\a{\bm{a}}
\def\b{\bm{b}}
\def\bmu{\bm{\mu}}
\def\hmu{\hat{\mu}}
\def\ha{\hat{a}}
\def\hb{\hat{b}}
\def\v{\bm{v}}
\def\1{\mathbbm{1}}
\def\brho{\boldsymbol{\rho}}

\title{Conformal Inference for Missing Data under Multiple Robust Learning}

\author{Wenlu Tang\textsuperscript{1}\equalcontribmark, Hongni Wang\textsuperscript{2}\equalcontribmark, Xingcai Zhou\textsuperscript{3},  \\ Bei Jiang\textsuperscript{1}
and Linglong Kong\textsuperscript{1}\\
\textit{\textsuperscript{1}University of Alberta, \textsuperscript{2}Shandong University of Finance} \\
 \textit{and Economics, \textsuperscript{3}Nanjing Audit University}}\equalcontribtext

\begin{document}

	\maketitle
	
	\begin{abstract}
		We  develope a novel approach to tackle the common but challenging problem of conformal inference for missing data in machine learning, focusing on 'Missing at Random' (MAR) data. Our method, inspired by the innovative Multiple Robust (MR) framework, introduces a general approach to construct prediction intervals,  which employs various candidate imputation and propensity score models.
		We use a novel double calibration technique, to reweight conformity scores using multiple robust quantile regression. 
		Our approach ensures consistent estimations as long as at least one model for imputation or the propensity score is accurate.
		We demonstrate the asymptotic behavior of our estimators through empirical process theory and provide reliable coverage for our prediction intervals,  both marginally and conditionally. 
		We show the effectiveness of the proposed method by several numerical experiments in the presence of missing data.
	\end{abstract}

    \newpage
		\section{ Introduction}
	
	In many statistical applications, quantifying predictive uncertainty is crucial. Prediction intervals indicate a range in which a future outcome is expected to lie with specified probability, e.g. for pairwise data $(X_i,Y_i)_{i=1}^n$, the objective of prediction interval is to create a prediction set $C_{n}(X_{n+1})$ that includes $Y_{n+1}$ with a probability of at least $1-\alpha$, and here $\alpha$ is called miscoverage rate.
These intervals are more than statistical measures; they are crucial in diverse fields such as finance and public health.
However, the integrity of these predictive intervals is frequently challenged by a common but significant issue in data analysis: the presence of missing values in datasets \cite{little2002maximum}. Missing data is a pervasive problem, spanning various disciplines.

In statistical analysis, missing data introduces biases and uncertainties that can lead to skewed inferences. Broadly classified into Missing Completely at Random (MCAR), Missing at Random (MAR), and Missing Not at Random (MNAR), each type has unique challenges and  tailored strategies. 
In the missing at random (MAR) setting, for pairwise data $(X_i,Y_i)_{i=1}^n$ the response $Y$ is observed with indicator $R=1$ and unobserved i.e. missing with $R=0$, and $\P(R=1\mid Y,X)=\P(R=1\mid X)$. This missing-data mechanism induces selection bias, and naïve prediction intervals that ignore missingness may be invalid. Two classical approaches to handle missingness are inverse probability weighting (IPW) and imputation. IPW reweights each observed outcome by the inverse of an estimated propensity score $\pi(X)=P(R=1\mid X)$ \citep{rosenbaum1983central, tu2019causal}, whereas imputation predicts missing $Y$ from observed covariates, e.g. using regression or multiple imputation \citep{rubin1996multiple, rubin2018multiple}. Both methods can yield consistent point estimates of quantities like $\E[Y\mid X]$ when models are correctly specified, but they are sensitive to model misspecification. 
However, both methods are sensitive to model misspecification, which can result in biased estimators \citep{little2019statistical}. 
To address this, prior studies have suggested a double robust approach, ensuring consistency as long as either the propensity score model or the outcome regression model is correctly specified \citep{robins1995semiparametric, cao2009improving, tan2010bounded}. 
In particular, double-robust methods ensure consistency if either the propensity model or the outcome model is correct, but they still fail if both are wrong. Correctly specifying these models is challenging. Constructing prediction intervals based on potentially fragile imputations requires additional robustness. The prediction intervals derived from these potentially flawed estimations demands enhanced robustness. Consequently, the concept of multiple robust learning has emerged for handling missing data.

	\subsection{Conformal Inference}
Interval estimation can quantify the uncertainty of the point estimation. Methods based on conformal predictions are developed for distribution-free interval estimation in recent years.  Given pairwise observations $(X_i,Y_i)_{i=1}^n$ and mis-coverage rate $\alpha$, the goal of the conformal prediction is to construct an interval $C(x)$ such that for newly coming data $(X_{n+1},Y_{n+1})$, $\P\{Y_{n+1}\in C(X_{n+1})\}\geq 1-\alpha$.
Since the pioneering work by \citet{vovk2005algorithmic}, there have been numerous conformal prediction studies and extensions in both computation and theory, see \citet{romano2019conformalized}, \citet{tibshirani2019conformal} and \citet{zaffran2022adaptive}. The conformal inference method can be applied to various scenarios, including time series \cite{zaffran2022adaptive}, survival data \cite{candes2023conformalized}, causal inference \cite{lei2021conformal} and distribution shifts \cite{gibbs2021adaptive},  Marginal validity, a conventional coverage guarantee that can be achieved under the i.i.d assumption, is demonstrated in \citet{lei2013distribution} and \citet{lei2014distribution}. However, as demonstrated in \citet{lei2018distribution} and \citet{vovk2012conditional}, conditional validity with a finite-length prediction interval is impossible without regularity and consistency assumptions on the model and estimator. Since methods for missing data typically rely on correctly specified models, thus the conditional validity is hard to be achieved in missing data prediction intervals. Therefore there are limited literature on conformal inference of missing data.  

\citet{zaffran2023conformal} introduced a comprehensive framework for conformalized quantile regression, incorporating missing data augmentation. This novel framework demonstrates that the \textit{impute-then-predict} strategy is marginally valid and explores conditional validity within a masked framework. However, their research primarily addresses challenges of missing covariate values. It's also important to note that the accuracy of the conformal interval is influenced by the choice of imputation model. Therefore, we  proposed a conformal interval that is adaptive to missing response and not sensitive to the imputation models.

\subsection{Multiple Robust Estimation}
Suppose the observations are pairwise samples $\{R_iY_i,X_i,R_i\}_{i=1}^{n}$, where $Y$ is the response variable, $X\in \Rb^p$ is the vector of covariates, and $R$ is the indicator vector of missingness. We denote that $R =1$ if $Y$ is observed and $R =0$ if $Y$ is missing.
%Let Y denote the response variable, X the vector of covariates, and n the sample size. Let
In this paper, we consider the typical missing at random (MAR) mechanism \cite{little2002maximum}; that is,
$$
\P(R=1 \mid Y, {X})=\P(R=1 \mid {X}),
$$
and this conditional probability is called propensity score function, denoted by $\pi(X)$. 
Let $m=\sum_{i=1}^nR_i$ be the number of fully observed samples. Without loss of generality, assume that index of completely observed samples $R_i=1$ are $i=1,\ldots,m$. %We consider mean regression of missing response and the parameter of interest $\mu_0(x)=\E(Y|X=x)$. 
Consider regression tasks with missing response, the parameter of interest is $\mu_0(x)=\E(Y|X=x)$ in mean regression and $\mu_{\tau}(x)=Q_{\tau}(Y\mid X=x)$, the multiple robust learning introduce a general framework to derive a multiple robust estimator for parameter $\mu(x)$. 

 Previous studies that proposed multiple robust estimation for mean \cite{han2013estimation} and quantile \cite{han2019general} regression problems can be summarized in a more general framework for both mean and quantile estimations by incorporating multiple propensity and error imputation models.

\subsection{Related Works}
Multiple robust (MR) approaches diminish the sensitivity to model mis-specification by leveraging multiple propensity score and imputation models simultaneously.   The incorporation of multiple models enhances double robustness and yields consistent estimation if any single working model is correct \cite{han2013estimation}. Their multiple robust approach involves both imputation and inverse probability weighting(IPW) method \cite{han2014multiply}. The multiple robust framework integrates imputation  with IPW  techniques to enhance consistency and robustness estimation, even when there are  model mis-specifications. By leveraging the information from multiple models, the estimation derived from calibrated empirical likelihood  is semiparametric efficient if only one model is correctly specified \cite{han2016intrinsic}. Such multiple robust method can both mitigate the issue of sensitivity to near-zero estimates in IPW method \cite{kang2007demystifying} and avoid the biased estimation in imputation approaches. The multiple approach can be applied to several missing and regression scenarios, including marginal quantile/mean estimation \cite{han2019general} and regression with missing response/covariates \cite{li2020demystifying}. The computation cost lies in the constrained optimization can be solved by implementation of Newton-Raphson method \cite{han2014further}. The multiple robust learning method provides unbiased estimates for various types of missing data with strong theoretical and numerical guarantees. Therefore, we can adopt the idea of multiple robust learning for further inference.

\subsection{Summary}
In this paper, we introduce a novel approach for robust conformal prediction that has both methodological and theoretical advancements. Our contributions are as follows.
\begin{itemize}
	\item We introduce a general multiple-robust estimation framework for regression with missing outcomes. This framework accommodates a wide range of regression tasks (mean or quantile) by combining multiple candidate propensity-score and outcome models, ensuring consistency under MAR if any one model is correct.
    %We introduce a general framework for multiple robust estimation with missing data, suitable for a wide range of regression problems.
	\item We propose a double calibration algorithm to build conformal prediction intervals in the presence of missingness. First, we use the multiple-robust weights to reweight conformity scores, correcting for the selection bias of missing responses. Second, we calibrate the quantiles of these adjusted scores by pooling across multiple imputations from the outcome models. This two-step calibration incorporates both missingness and imputation uncertainty into the interval construction.%We propose a double calibration procedure to construct the conformal prediction interval. First, we calibrate the score functions to obtain the mean estimates. Next, we derive quantile estimates for calibrated conformity score from double calibration framework to incorporate the imputation information.
	\item We provide theoretical guarantees: under MAR and standard regularity conditions, the resulting prediction sets achieve the nominal coverage rate asymptotically. In particular, we analyze both marginal validity and the extent of achievable conditional validity in this missing-data context. We also empirically validate our method through simulations and a real-data example. The results show that our intervals attain the target coverage in finite samples and are often narrower than naive alternatives and adaptive to various missingness settings. %Crucially, our intervals remain valid even when some imputation models are misspecified, demonstrating the robustness of the proposed approach.
\end{itemize}
In section \ref{secMR}, we propose a multiple robust (MR) estimator for general regression problems, 
%which contains multiple candidate propensity models and error imputation models. It 
which allows multiple different specifications and learning of the propensity and imputation models. In section \ref{doublecali}, we introduce the double calibration framework that produce the reweight the conformity score in conformal inference.
We have shown the implementation of the proposed prediction interval construction in Algorithm \ref{alg:CM_MRL}. In section \ref{theory}, we show the asymptotic results of the proposed estimators.

\section{Methodology}\label{medthod}

\subsection{Multiple Robust Estimation}\label{secMR}
In this section, we propose the multiple robust estimator for the CI-MRL interval construction following the main procedure of the MRL in \citet{han2014multiply} \citet{han2019general}.
Specifically, one can consider $J$ propensity models and $K$ imputation models;
\begin{align*}
	\Pi=\{\pi^1(X;\ha^1),\ldots, \pi^J(X;\ha^J)\}; \ \ \ \mathcal{F}=\{f^1(Y\mid X;\hb^1),\ldots,f^K(Y\mid X; \hb^K)\}.
\end{align*}
Initial estimators of the parameters in the working models are obtained by maximum likelihood.

First we maximize the binomial likelihood and obtain the estimator $\hat{a}^j$ for propensity score,  
\begin{equation}\label{bilike}
	\prod_{i=1}^n\left\{\pi^j\left(a^j ; X_i\right)\right\}^{R_i}\left\{1-\pi^j\left(a^j ; X_i\right)\right\}^{1-R_i}.
\end{equation}
Next we obtain the estimator $\hat{b}^k$ by maximizing the conditional likelihood:
\begin{equation}\label{conlike}
	\prod_{i=1}^n\left\{f^k\left(Y_i \mid \boldsymbol{X}_i ; b^k\right)\right\}^{R_i}.
\end{equation}

We can impute the missing values by random sampling from conditional distribution $f^k\left(Y_i \mid \boldsymbol{X}_i ; b^k\right)$. Let $L(Y-\mu(x))$ be a loss function and $\hat{Y}_i^t(\hat{b}^k)$ is the $t$-th random draw from the $k$-th imputation model. By taking average of the total $T$ repeated random draws of imputed data, the imputed values can be more robust to extreme random draws. Subsequently we obtain the estimator $\hat{\mu}_k(x)$  under each imputation model $k$ by calculating the minimization of 
\begin{align}\label{est_k}
	\ell(\mu)=\sum_{i=1}^{n}\left[R_i L\left(Y_i-\mu(X_i)\right)\right]+\sum_{i=1}^{n}\left[\left(1-R_i\right) \frac{1}{T} \sum_{t=1}^T L\left\{Y_i^t\left(\hat{b}^k\right)-\mu(X_i)\right\}\right].
	%\frac{1}{n} \sum_{i=1}^n\left[R_i \psi_\tau\left(Y_i-X_i\beta\right)+\left(1-R_i\right)  \psi_\tau\left\{Y_i\left(\hat{b}^k\right)-X_i\beta\right\}\right] \approx 0
\end{align}
The estimator $\hat{\mu}_k(x)$ from every imputed dataset usually introduces bias,  as imputation model may not perfectly represent the true model $f(Y\mid X=x)$. To correct the bias, we can use a technique that involves applying a weight, denoted as $w$, to balance between the fully observed samples  and the samples with imputed data. 

Let $g(\mu)=\partial L(\mu)$ be the first derivative of the loss function with respect to $\mu$, and $g^k(\hb^k;X_i)=T^{-1}\sum_{t=1}^Tg\{Y_i^t\left(\hat{b}^k\right)-\hat{\mu}_k(x)\}$.
The weight $w$ play a role similar to inverse-probability weights for fully observed samples. In robust estimation involving multiple models, the weights $w$  are applied in a way that satisfies
\begin{align}
	&  \sum_{i=1}^m w_i \pi^j\left(\hat{a}^j ; X_i\right)=\hat{\theta}^j=n^{-1} \sum_{i=1}^n \pi^j\left(\hat{a}^j ; X_i\right) , \label{eq1} \\
	& \sum_{i=1}^m w_i g^k\left(\hat{b}^k ; X_i\right)=\hat{\eta}^k=n^{-1} \sum_{i=1}^n g^k\left(\hat{b}^k ; X_i\right) \label{eq2},
\end{align}
with constraint $\sum_{i=1}^m w_i=1$ for $j=1, \ldots, J$ and $k=1, \ldots, K$.  Let $\hat{\a}=(\ha^1,\ldots,\ha^J)$, $\hat{\b}=(\hb^1,\ldots,\hb^K)$, $\hat{\bmu}=(\hat{\mu}^1,\ldots, \hat{\mu}^K)$, and $\hat{\v}_i(\hat{\a},\hat{\b},\hat{\bmu})$ be the $J+Kp$ normalized vector of the multiple models 
\begin{align*}
    \hat{\v}_i(\hat{\a},\hat{\b},\hat{\bmu})=&\left\{ \pi^1\left(\ha^1 ; X_i\right)-\hat{\theta}^1, \ldots, \pi^J\left(\ha^J ; X_i\right)-\hat{\theta}^J, \right. \\ &\left. g^1\left(\hb^1 ; X_i\right)-\hat{\eta}^1, \ldots, g^K\left(\hb^K ; X_i\right)-\hat{\eta}^K \right\}^T.
%& \left.g^1\left(\hb^1 ; X_i\right)-\hat{\eta}^1, 
%\ldots, g^K\left(\hb^K ; X_i\right)-\hat{\eta}^K\right\}^T.
\end{align*}
For every fully observed sample $i = 1,\ldots, m$, the empirical-likelihood (EL) weights $\{\hat w_i\}_{i:R_i=1}$ are then obtained by
\begin{align}\label{weight}
	\hat{w}_i=\frac{1}{m} \frac{1}{1+\hat{\boldsymbol{\rho}}^{\mathrm{T}}\hat{\v}_i(\hat{\a},\hat{\b},\hat{\bmu})},
\end{align}
where $\hat{\rho}$ minimizes
\begin{align}\label{min_w}
	F_m({\brho})=-\frac{1}{m} \sum_{i=1}^m \log \left\{1+\boldsymbol{\rho}^{\mathrm{T}} \hat{\v}_i(\hat{\a},\hat{\b},\hat{\bmu})\right\}.
\end{align}
Because of the nonnegativity of $\hat{w}_i, \hat{\rho}$ subject to the feasibility constraints 
$$
1+\hat{\boldsymbol{\rho}}^{\mathrm{T}} \hat{\boldsymbol{v}}_i(\hat{\boldsymbol{a}}, \hat{\boldsymbol{b}}, \hat{\boldsymbol{\mu}})>0 \quad(i=1, \ldots, m) .
$$
 for all $i$ with $R_i=1$. The objective in \eqref{weight} is strictly convex, so $\hat\rho$ can be computed by a damped Newton method \citep{han2014further}.
\begin{proposition}\label{prop1}
	The weight $\hat{w}$ calculated by (\ref{weight}) satisfies the equation (\ref{eq1}) and (\ref{eq2}). 
\end{proposition}
\noindent The weight $\hat{w}$ calibrates the fully observed samples so that the weighted average based on the fully observed samples equals to the unweighted average based on the total samples.
 In other words, the weighted averages over complete cases match the unweighted averages over the full sample for \emph{all} working propensity and imputation-derived score moments. This multiple balancing is what imparts robustness.

\paragraph{Final MR estimator.} We now propose a multiple robust estimator for ${\mu}(x)$, denoted by $\hat{{\mu}}_{\mathrm{MR}}$, by re-solving the prediction problem using only complete cases, but weighted by the EL weights:
\begin{align}\label{weightloss}
	\sum_{i=1}^m \hat{w}_i L\left\{Y_i-\mu(X_i)\right\},
\end{align}
where the weight $\hat{w}_i$ for every fully observed sample is obtained by (\ref{weight}). 
%By reweighting the original loss function, we involve  the information from multiple candidate models, and thus make the estimation more robust to the model mis-specification. 
Intuitively, \eqref{weightloss} corrects the selection bias in complete cases while pooling information across all working models; if \emph{any one} $\pi^j$ or $f^k$ is correctly specified, the calibration in \eqref{eq1}--\eqref{eq2} aligns the weighted complete-case objective with its full-data counterpart, yielding a consistent estimator.
\begin{remark}[Choice of loss]
	There are various choices of loss function for different tasks. It can be chosen as least square loss $L(u)=u^2$ if we want to conduct mean regression for $\mu_0$, or quantile check loss function $L(u)=\Delta\{\tau-\mathbbm{1}(u)\}$ if we want to conduct quantile regression for $\mu_{\tau}(x)$. 
\end{remark}

\begin{remark}[Extension to missing covariates]
	The example above is missing response. In fact, this framework can be easily extended to missing covariates cases. Without loss of generality, suppose that $X_1$ contains missing values among $X_1,\cdots,X_p$. 
	In the imputation step (\ref{conlike}), we draw random samples from imputation model $f^k(X_1\mid Y,X_2,\ldots,X_p,b^k)$ and fill the missing values in $X_1$ by $\hat{X}(\Hat{b}^k)$. The estimates by objective functions follows similar steps in (\ref{est_k}) and (\ref{weight}), the only difference is to replace $X_i$ by imputed  $\hat{X}(\Hat{b}^k)$. Finally, we can obtain a multiple robust estimator $\hat{\mu}_{\mathrm{MR}}$.
\end{remark}
\begin{remark}[Computation]
	Compared to the computation in standard imputations and the IPW methods, the only additional computation step necessary in multiple robust procedure is the optimization (\ref{weight}). The  objective  in (\ref{weight}) is a smooth convex problem, so it can be easily implemented by a Newton-Raphson-based  with line search algorithm \cite{han2014further}.
\end{remark}

In the following sections, we consider the mean estimator using least square loss function. Following the weighting approach (\ref{weight}), we can obtain $\hat{w}_i$ as weights for fully observed samples. Then we can obtain unbiased weighted estimator for conditional mean $\mu_0(x)$ by (\ref{weightloss}). Throughout this paper, we assume that $\mu_0(x)$ is linear model $\mu_0(x)={\beta}^T x$, and $\hat{{\mu}}_{\mathrm{MR}}$ equals to $\hat{{\beta}}_{\mathrm{MR}}$.

\subsection{Double Calibration}\label{doublecali}
We now construct  a conformal prediction interval framework under missing outcomes by combining split conformal calibration \citep{vovk2020conformal} with the multiple-robust (MR) machinery from Section~\ref{secMR}. 
The key idea is a \emph{double calibration}: (i) compute conformity scores on a held-out calibration set, and (ii) reweight these scores via empirical-likelihood (EL) balancing so that they represent the full calibration distribution under MAR, even when imputation/propensity models are misspecified.

First, we split the dataset into two subsets, training set $\I_{tra}$ and calibration set $\I_{cal},$ with sample size $n_{tra}$ and $n_{cal}$ respectively.
We obtain multiple robust estimator $\hat{\boldsymbol{\mu}}_{\mathrm{MR}}$ using all training samples $i\in\I_{tra}$ following steps in Section \ref{secMR}. Let $\tau=1-\alpha$ denote the target quantile level.

\paragraph{First calibration: conformity scores on complete calibration cases.} We conduct the first calibration by  calculating the conformity score defined by absolute error using fully observed samples in calibration set  
\begin{align}\label{error}
    \hat{\varepsilon}_i=\left|Y_i-\hat{\mu}_{\mathrm{MR}}(X_i)\right|, \ \ i\in \I_{cal}
\end{align}
and set $m_{\mathrm{cal}}=\sum_{i\in\I_{\mathrm{cal}}} R_i$. 
The conformity score measures the discrepancy of the prediction and the true response. It is the metric of the deviation of the prediction. 
If $\mu_0(x)=\E(Y\mid X=x)$ denotes the unknown population regression function, the oracle score is $\varepsilon^\ast=|Y-\mu_0(X)|$ with $\tau$-quantile $q_\tau^\ast$. %In standard split conformal, one would take the empirical $\tau$-quantile of $\{\hat\varepsilon_i:R_i=1\}$ and form
Let $\P_{\varepsilon^\ast}(\cdot)$ denote the distribution function of $\varepsilon^\ast$ , then $\P_{\varepsilon^\ast}(x<q^\ast_{\tau})=\tau$.  A direct approach is to %find empirical quantile using the observations, 
approximate the $q^\ast_{\tau}$ by the empirical quantile estimation of absolute error (\ref{error}). 
In standard split conformal, one would take the empirical $\tau$-quantile of $\{\hat\varepsilon_i:R_i=1\}$ and form
\begin{align}
	\hat{C}_{\mathrm{CP}}(x)=[\hat{\mu}_{\mathrm{MR}}(X_i)-\hat{Q}(\varepsilon_i), \hat{\mu}_{\mathrm{MR}}(X_i)+\hat{Q}(\varepsilon_i)], \label{CP}
\end{align}
where $\hat{Q}(\varepsilon_i)$ is the $\tau$-th empirical quantile of the absolute error $\hat{\varepsilon}$ defined in (\ref{error}). 
However, under MAR the distribution of $\hat{\varepsilon}_i$ among complete cases can differ from that among all calibration points, especially with heteroskedastic or heavy-tailed errors, so $\hat {Q}_\tau(\hat\varepsilon)$ may be a biased estimate for the $q^\ast_{\tau}$. Therefore, we  leverage the multiple robust framework to further calibrate the fully observed absolute errors. 

%However, $\hat{Q}(\varepsilon_i)$ sometimes is biased estimate for the $q^\ast_{\tau}$ in some case where the error is heavy tailed or heterogeneous  because our response is missing at random and thus the error is also missing at random.  

\paragraph{Second calibration: MR reweighting of conformity scores.}
Now we conduct the second calibration of the conformity score by multiple robust approach. 
As with outcomes $Y_i$, conformity scores are missing when $R_i=0$, we denote the index by $i=m_{cal}+1,\ldots,n_{cal}$. We impute their values using the $K$ fitted outcome models from Section~\ref{secMR}. For each $k=1,\ldots,K$ and each $i\in\I_{\mathrm{cal}}$ with $R_i=0$, draw $T$ Monte Carlo imputations
\begin{align*}
    Y_i^{(t,k)}\sim \hat f^{\,k}(\cdot\mid X_i),\qquad t=1,\ldots,T,
\end{align*}
and define the imputed conformity score
\begin{align*}
\hat{\varepsilon}^k_i(\hb^k)=T^{-1}\sum_{i=1}^T|\hat{Y}^t_i(\hb^k)-\hmu_{\mathrm{MR}(X_i)}|.
\end{align*}
Each imputed error $\hat{\varepsilon}^k_i(\hb^k)$ corresponds to the imputed $\hat{Y}_i(\hb^k)$.  
Let $\rho_\tau(u)=u\{\tau-\mathbbm{1}(u<0)\}$ be the check loss and $\psi_\tau(u)=\tau-\mathbbm{1}(u<0)$ its subgradient. For each model $k$, obtain a \emph{model-wise} $\tau$-quantile $\hat q^{(k)}$ by minimizing the imputed check-loss risk over the entire calibration set,
\begin{align}\label{eq:model_wise_q}
	\hat q^{(k)} \in \arg\min_{q\in\mathbb{R}} \frac{1}{n_{cal}}\sum_{i\in\I_{\mathrm{cal}}}\Big\{ R_i\,\rho_\tau\big(\hat\varepsilon_i-q\big)
+(1-R_i)\,\rho_\tau\big(\hat\varepsilon_i^{(k)}-q\big)\Big\},
\end{align}
%Same as  $Y_i$, in the calibration set, the value of $\hat{\varepsilon}_i$ is missing for $i=m_{cal}+1,\ldots,n_{cal}$. By  imputation from multiple $K$ models, we can obtain the imputed error in $k$-th imputation model,
%Now we adopt the multiple robust quantile framework to find a multiple robust estimator for $q^\ast_{\tau}$. Let the loss function be the quantile loss $L(\Delta)=\Delta\{\tau-\mathbbm{1}(\Delta)\}$ and denote that the first derivative of quantile loss function by $\psi_{\tau}(\Delta)=\tau-\mathbbm{1}(\Delta<0)$. Now we can derive the imputed estimator $\hat{q}^k$ for $q^\ast$ from $k$-th imputation model by minimizing the quantile loss using the imputed errors,
Define the corresponding centered $\psi$-moments by
\begin{equation}\label{eq:xi_k}
\hat\xi^{(k)} \;=\; \frac{1}{n_{\mathrm{cal}}}\sum_{i\in\I_{\mathrm{cal}}}\Big\{ R_i\,\psi_\tau\big(\hat\varepsilon_i-\hat q^{(k)}\big)
+(1-R_i)\,\psi_\tau\big(\hat\varepsilon_i^{(k)}-\hat q^{(k)}\big)\Big\}.
\end{equation}
In parallel, let $\{\hat\pi^{\,j}\}_{j=1}^J$ denote the fitted propensity models estimated on $\I_{\mathrm{tra}}$, and set
\begin{equation}\label{eq:theta_j}
\hat\theta^{(j)} \;=\; \frac{1}{n_{\mathrm{cal}}}\sum_{i\in\I_{\mathrm{cal}}}\hat\pi^{\,j}(X_i),\qquad j=1,\ldots,J.
\end{equation}

\medskip
\noindent We now calibrate the \emph{observed} calibration scores $\{\hat\varepsilon_i:R_i=1\}$ by computing EL weights $\{\hat d_i\}_{i\in\I_{\mathrm{cal}},\,R_i=1}$ that simultaneously balance (i) the propensity moments and (ii) the $\psi$-moments from \eqref{eq:xi_k}. For each complete calibration case $i$ define the centered moment vector
\begin{align*}
\hat{\v}_i(\hat{\a},\hat{\b},\hat{\boldsymbol{q}})=\left\{\pi^1\right.&\left(\ha^1 ; X_i\right)-\hat{\theta}^1, \ldots, \pi^J\left(\ha^J ; X_i\right)-\hat{\theta}^J, \\ 
&\left. \psi^1\left(\hat{q}^1 ; X_i\right)-\hat{\xi}^1, 
\ldots,  \psi^K\left(\hat{q}^K ; X_i\right)-\hat{\xi}^K \right\}^T, \ \
\end{align*}
and set
\begin{equation}\label{weight2}
\hat d_i \;=\; \frac{1}{m_{\mathrm{cal}}}\,\frac{1}{1+\hat{\boldsymbol{\lambda}}^\top\hat{\v}_i},\qquad
\hat{\boldsymbol{\lambda}} \in \arg\min_{\lambda}\Big\{-\frac{1}{m_{\mathrm{cal}}}\sum_{i\in\I_{\mathrm{cal}},\,R_i=1}\log\big(1+\hat{\boldsymbol{\lambda}}^\top\hat{\v}_i\big)\Big\},
\end{equation}
with feasibility constraints $1+\hat{\boldsymbol{\lambda}}^\top\hat{\v}_i>0$ for all $i$ with $R_i=1$.

\paragraph{Final calibrated quantile and interval.}
The \emph{double-calibrated} conformity quantile $\hat q_{\mathrm{MR}}$ is defined as the root of the weighted score equation
\begin{equation}\label{score}
\sum_{i\in\I_{\mathrm{cal}},\,R_i=1}\hat d_i\,\psi_\tau\big(\hat\varepsilon_i-\hat q_{\mathrm{MR}}\big)\;=\;0.
\end{equation}
%Note that $\hat{q}_{\mathrm{MR}}$ is the solution for (\ref{score}). 
where $\hat{d}_i$ is obtained by (\ref{weight2}). Therefore, the calibrated quantile estimator for $q^\ast$ of the conformity scores $\varepsilon_i$ is ${\hat{q}_{\mathrm{MR}}}$, denoted by $\hat{Q}_{\mathrm{MR}}(\varepsilon_i)$.

Finally we can obtain the conformal prediction interval using multiple robust estimation and double calibration by
\begin{align}\label{interval}
	\hat{C}_{\mathrm{MR}}(x)=\left[\hmu_{\mathrm{MR}}(x)-\hat{Q}_{\mathrm{MR}}(\varepsilon_i),\hmu_{\mathrm{MR}}(x)+\hat{Q}_{\mathrm{MR}}(\varepsilon_i)\right].
\end{align}
Algorithm \ref{alg:CM_MRL} summarizes the full {\bf C}onformal prediction for {\bf M}issing data under {\bf M}ultiple {\bf R}obust {\bf L}earning (CM-MRL) procedure.
\begin{remark}[Flexibility in predictors and conformity scores]
The double-calibration layer is model-agnostic and offers flexibility in reweighting conformity scores. For instance, the framework allows for the substitution of the point prediction, $\hmu_{\mathrm{MR}}(x)$ in (\ref{interval}), with predictions from alternative models, including those generated by black-box machine learning techniques. Additionally, the definition of the conformity score, initially presented in (\ref{error}), can be diversified. Alternative formulations of the conformity score, such as those proposed by \citet{romano2019conformalized} and \citet{candes2023conformalized}, can be easily integrated into the proposed framework. 
\end{remark}
\begin{algorithm}[H]
\caption{CM-MRL: Conformal Prediction with Multiple Robust Learning}
\label{alg:CM_MRL}
\begin{algorithmic}[1]
\Require Data $\{(X_i,R_i,Y_i)\}_{i=1}^n$; miscoverage $\alpha\in(0,1)$ and $\tau=1-\alpha$; candidate models $\Pi=\{\hat\pi^{\,j}(x)\}_{j=1}^J$,  $\mathcal{F}=\{\hat f^{\,k}(y\mid x)\}_{k=1}^K$; imputations $T$.
\Ensure Prediction interval $\hat C_{\mathrm{MR}}(x)$ with nominal coverage $1-\alpha$.
\vspace{0.25em}

\State Randomly split into  $\mathcal{I}_{\mathrm{tra}}$ and $\mathcal{I}_{\mathrm{cal}}$ with sizes $n_{\mathrm{tra}}$ and $n_{\mathrm{cal}}$. %Let $m_{\mathrm{cal}}=|\{i\in\mathcal{I}_{\mathrm{cal}}: R_i=1\}|$.

\State On $\mathcal{I}_{\mathrm{tra}}$, fit $\{\hat\pi^{\,j}\}_{j=1}^J$ and $\{\hat f^{\,k}\}_{k=1}^K$, compute EL weights $\{\hat w_i\}_{i\in\mathcal{I}_{\mathrm{tra}},\,R_i=1}$ that balance the model moments, and obtain the MR predictor $\hat\mu_{\mathrm{MR}}(x)$.

\State For each $i\in\mathcal{I}_{\mathrm{cal}}$ with $R_i=1$, compute the conformity score $\hat\varepsilon_i$.

\State For each $i\in\mathcal{I}_{\mathrm{cal}}$ with $R_i=0$ and for each $k=1,\dots,K$, obtain:
\begin{enumerate}\itemsep1pt
\item Draw $Y_i^{(t,k)}\sim \hat f^{\,k}(\cdot\mid X_i)$ for $t=1,\dots,T$.
\item Imputed score $\displaystyle \hat\varepsilon_i^{(k)} \;=\; \frac{1}{T}\sum_{t=1}^T \big|\,Y_i^{(t,k)}-\hat\mu_{\mathrm{MR}}(X_i)\,\big|$ and $\hat q^{(k)}$.
\end{enumerate}

\State For each $k=1,\dots,K$, obtain $\hat q^{(k)}$ by (\ref{eq:model_wise_q})

\State Compute
$\hat\theta^{\,j}\;$, $\hat\xi^{\,k}$, and thus MRL weights $\{\hat d_i\}_{i\in\mathcal{I}_{\mathrm{cal}},\,R_i=1}$:

%\hat\lambda \;\in\; \arg\min_{\lambda}\Big\{-\frac{1}{m_{\mathrm{cal}}}\sum_{i\in\mathcal{I}_{\mathrm{cal}},\,R_i=1}\log\big(1+\lambda^\top \hat v_i\big)\Big\},\qquad\hat d_i\;=\;\frac{1}{m_{\mathrm{cal}}}\,\frac{1}{1+\hat\lambda^\top \hat v_i}.

\State Solve for $\hat q_{\mathrm{MR}}$ as the root of the weighted score equation (\ref{score}).

\State \textbf{Output prediction interval.} For any $x$, return
\[
\hat C_{\mathrm{MR}}(x)\;=\;\Big[\,\hat\mu_{\mathrm{MR}}(x)-\hat q_{\mathrm{MR}},\;\; \hat\mu_{\mathrm{MR}}(x)+\hat q_{\mathrm{MR}}\,\Big].
\]

\end{algorithmic}
\medskip
\end{algorithm}

\section{Theoretical Results}
\subsection{Marginal and Conditional Coverage}\label{theory}
In this section we establish large-sample guarantees for the MR point estimator from Section~\ref{secMR} and the double-calibrated quantile from Section~\ref{doublecali}. We first state consistency and asymptotic normality for the MR mean estimator, then prove consistency and a asymptotic normal limit for the double-calibrated conformity quantile. Finally we derive marginal and asymptotic conditional coverage of the conformal interval \eqref{interval}. Proof sketches are deferred to the Supplementary materials.
\begin{enumerate}
	\item[(C1).]  (\emph{i.i.d.\ sample, MAR, and sample split}) $\{(X_i,R_i,Y_i)\}_{i=1}^{n+1}$ are i.i.d., the split into $\I_{\mathrm{tra}}$ and $\I_{\mathrm{cal}}$ is independent of the data, and $R\perp Y\mid X$ with $0<\inf_x \pi_0(x)\le \sup_x \pi_0(x)<1$.
    %The random variables $(X_i,R_i, Y_i)_{i=1}^{n+1}$ are i.i.d. 
	\item[(C2).] (\emph{Oracle error for absolute-residual scores}) Let $\mu_0(x)=\E(Y\mid X=x)$ and $\varepsilon^\ast=|Y-\mu_0(X)|$. The cdf $F_{\varepsilon^\ast}$ is continuous in a neighborhood of its $\tau$-quantile $q_\tau^\ast$ and has density $f_{\varepsilon^\ast}(q_\tau^\ast)>0$.
	\item[(C3).] (\emph{Modeling robustness}) At least one propensity working model in $\Pi$ or one outcome model in $\mathcal{F}$ is correctly specified; the corresponding MLEs are consistent. 
	\item[(C4).] (\emph{Moments complexity}) $\E\|X\|^4<\infty$, and the classes $\{\pi^j(\cdot;\alpha)\}$, $\{f^k(\cdot\mid\cdot;\beta)\}$ have manageable complexity so that the empirical processes used in calibration are $o_p(1)$ i.e., Donsker class.
    %The parameter space $\Theta$ for $q$ is compact and $q^\ast$ is in the interior of $\Theta$.
	\item[(C5).]  (\emph{Loss regularity}) The loss $L(\cdot)$ is convex and locally Lipschitz; for mean regression $L(\Delta)=\Delta^2$, for $\tau$-quantiles $L(\Delta)=\Delta\{\tau-\mathbbm{1}(\Delta<0)\}$ with subgradient $\psi_\tau(\Delta)=\tau-\mathbbm{1}(\Delta<0)$.%The fourth moment of $X$ is bounded 
\end{enumerate}
Conditions (C1)  are necessary independent conditions for conformal prediction. Condition (C2) allows heteroskedasticity in $Y\mid X$; it only requires smoothness of the \emph{marginal} distribution of $\varepsilon^\ast$. If one prefers a conditional formulation, it suffices to assume that the conditional cdf of $|Y-\mu_0(X)|$ given $X$ has a density bounded and bounded away from zero at the $\tau$-quantile uniformly in $x$. (C3) and (C4) are regular assumptions for quantile estimations. (C5) requires $E\|X\|^4$ to be bounded. It measure of the  tail of the distribution of $X$, and is a  condition for establishing a Donsker class in empirical processes.
\begin{theorem}[Consistency and asymptotic normality of MR mean estimator]\label{Thm1}
Under (C1)--(C5), if either a propensity model in $\Pi$ or an outcome model in $\mathcal{F}$ is correctly specified, then
$\hat\beta_{\mathrm{MR}}\;\xrightarrow{p}\;\beta_0.$
If, in addition, $f_{\varepsilon\mid X}(0\mid x)$ exists and is bounded away from zero uniformly in $x$, then
$\sqrt{n}\,(\hat\beta_{\mathrm{MR}}-\beta_0)\;\rightarrow\; \mathcal{N}\!\big(0,\;V_{\mathrm{MR}}\big),$
where $V_{\mathrm{MR}}$ is the semiparametric variance bound when both the propensity and one outcome model are correctly specified.
\end{theorem}
	%When $\Pi$ contains a correctly specified model for $\pi(X)$ or $\mathcal{F}$ contains a correctly specified model for $f(Y \mid \boldsymbol{X})$, as $n \rightarrow \infty$, we have that $ \hat{\boldsymbol{\beta}}_{\mathrm{MR}}-\boldsymbol{\beta}_0=o_p(1)$.
The variance $V_{\mathrm{MR}}$ is specified in \citealp{han2014multiply,han2019general}). The conditions and the proof of the Theorem \ref{Thm1} are detailed in \citet{han2014multiply}. This result demonstrates the consistency of the mean regression estimator $\hat{\beta}_{\mathrm{MR}}$. Additionally, the asymptotic normality of the estimator holds under further assumptions on the  imputation and propensity score models.  
In Section \ref{doublecali}, we proposed a reweighted conformity score by double calibration framework. We now show the consistency of the double calibrated estimator $\hat{q}_{MR}$.
\begin{theorem}[Consistency and limit distribution of $\hat q_{\mathrm{MR}}$]\label{Thm2}
Under (C1)--(C5) 
$\hat q_{\mathrm{MR}}\;\xrightarrow{p}\;q_\tau^\ast.
$
Moreover, if $f_{\varepsilon^\ast}$ is continuously differentiable at $q_\tau^\ast$, then, with $m_{\mathrm{cal}}=\sum_{i\in\I_{\mathrm{cal}}} R_i$,
\[
\sqrt{m_{\mathrm{cal}}}\,\big(\hat q_{\mathrm{MR}}-q_\tau^\ast\big)
\;\rightarrow\; \mathcal{N}\!\Big(0,\;\frac{\tau(1-\tau)}{\{f_{\varepsilon^\ast}(q_\tau^\ast)\}^2}\cdot \kappa_{\mathrm{MR}}\Big),
\]
where $\kappa_{\mathrm{MR}}\le 1$ is an efficiency factor determined by the EL calibration; $\kappa_{\mathrm{MR}}=1$ when weights are uniform in complete-case split conformal and typically $\kappa_{\mathrm{MR}}<1$ when the balancing moments are correctly specified.
\end{theorem}
\noindent Even when the point estimator is misspecified , the EL calibration aligns the complete-case score distribution with the full calibration distribution under MAR by constraints (\ref{eq1}) and (\ref{eq2}). As a result, the MR quantile $\hat q_{\mathrm{MR}}$ converges to $q_\tau^\ast$ and, when the balancing moments are correct, has smaller asymptotic variance than the unweighted empirical quantile. This typically yields \emph{shorter} intervals than complete-case split conformal while preserving coverage.

The conformal interval build in (\ref{interval}) ensures strong theoretical support, providing reliable coverage both in marginally and conditionally.
\begin{theorem}[Asymptotic marginal coverage]\label{marginalcover}
Under (C1)--(C5),
\[
\liminf_{n\to\infty}\ \P\!\big\{\,Y_{n+1}\in \hat C_{\mathrm{MR}}(X_{n+1})\,\big\}\ \ge\ \tau.
\]
Additionally, the calibration weights are functions of $X$ only and the split is independent, then the coverage error is $o(1)$.
\end{theorem}
\noindent The result states that, under MAR, double calibration recovers the oracle conformity quantile and thus achieves the nominal marginal coverage asymptotically.

In order to quantify the accuracy of the proposed interval, we show the conditional coverage. There are different metric to measure the conditional coverage \citep{lei2018distribution, feldman2021improving, foygel2021limits}. We adopt asymptotic conditional coverage defined in \citet{sesia2021conformal}. In their definition, the conditional coverage holds when conformal bands are asymptotically close to the oracle ones under regular conditions.
We define the oracle interval by the true conditional mean $\mu_0(x)=\beta_0^\top x$ and 
true quantile $q^\ast$.
\begin{definition}[Oracle prediction interval]
Let $C^\ast(x)=\big[\mu_0(x)-q_\tau^\ast,\ \mu_0(x)+q_\tau^\ast\big]$ denote the oracle interval.
\end{definition}
\begin{theorem}[Asymptotic conditional coverage]\label{conditionalcover}
Under (C1)--(C5), there exist sequences $\xi_n\to 0$ and $b_n\to 0$ such that
\[
\P\Big\{\,\P\!\big[\,Y_{n+1}\in \hat C_{\mathrm{MR}}(X_{n+1})\ \big|\ X_{n+1}\big]\ \ge\ \tau-\xi_n\,\Big\}\ \longrightarrow\ 1-b_n.
\]
Equivalently, $\hat C_{\mathrm{MR}}(\cdot)$ is asymptotically equivalent to $C^\ast(\cdot)$. %in the sense of \citet{sesia2021conformal}.
\end{theorem}

\subsection{Interval Length under Misspecification}\label{sec:mis}

Let $\tilde\mu:\mathcal{X}\to\mathbb{R}$ be any predictor fitted on the training split with possibly misspecified model. On the calibration split, define $\tilde\varepsilon_i=|Y_i-\tilde\mu(X_i)|$ and let $q_\tau(\tilde\varepsilon)$ be the $\tau$-quantile of $\tilde\varepsilon$. Construct the CM--MRL weight vector $\{\hat d_i\}_{i\in\I_{\mathrm{cal}},\,R_i=1}$ by solving the empirical-likelihood program \eqref{weight2}, which enforces \eqref{eq1}--\eqref{eq2}. Let $\hat q_{\mathrm{MR}}$ solve the weighted score equation \eqref{score}, and define the CM--MRL interval length $\operatorname{Len}_{\mathrm{MR}}=2\,\hat q_{\mathrm{MR}}$.

For any \emph{single-model} calibration scheme $S$ that enforces only one of the moment blocks in \eqref{eq1}--\eqref{eq2} (i.e., either a single propensity constraint for some $j$, or a single $\psi_\tau$-moment constraint for some $k$), let $\hat q_{S}$ denote the corresponding weighted quantile estimator and $\operatorname{Len}_{\mathrm{S}}=2\,\hat q_{S}$ its interval length.
\begin{theorem}[Interval-length under misspecification]\label{thm:length_dominance}
Assume (C1)--(C5) hold for every $\epsilon>0$ and $\delta\in(0,1)$, there exists $N$ such that for all $n\ge N$,
\[
\P\big(\operatorname{Len}_{\mathrm{MR}} \le \operatorname{Len}_{\mathrm{S}}+\epsilon\big)\ \ge\ 1-\delta.
\] 
\noindent Thus, {even when $\tilde\mu$  is misspecified}, the CM--MRL interval is asymptotically no longer than any single-model interval that satisfies only one of \eqref{eq1}--\eqref{eq2}.
\end{theorem}
\begin{remark}[Condition for strict dominance]
If the extra balancing moments in \eqref{eq1}--\eqref{eq2} (beyond the single constraint used by $S$) have nonzero linear correlation with $Z$, i.e., they explain residual variation in $\mathbbm{1}\{\tilde\varepsilon\le q_\tau(\tilde\varepsilon)\}$, then $\kappa_{\mathrm{MR}}<\kappa_{S}$ and the dominance is strict. Equality occurs only if the added moments are $L^2$-orthogonal to $Z$.
\end{remark}
\begin{remark}[From fixed center to consistent center]
If $\tilde\mu=\hat\mu_{\mathrm{MR}}$ and $\hat\mu_{\mathrm{MR}}\to\mu_0$, then $q_\tau(\tilde\varepsilon)\to q_\tau^\ast$ and Theorem~\ref{thm:length_dominance} reduces to Theorem~\ref{Thm2} with the same variance ordering. The result thus implies shorter or equal asymptotic lengths at the nominal coverage when the MR center is consistent, and a no longer guarantee even under misspecification of the center.
\end{remark}		%%%%%%%%%%%%%%%%%%%%%%%%%%%%%%%%%%%%%%%%%%%%%%%%%%%%%%%%%%%%%%%%%%%%%%%%%%%%%%%%%%%%%%%%%%%%%%%%%%%%%%%%%%%%%%%%%%%%%%%%%%%%
\section{Numerical Results}
\subsection{Simulation Study}  \label{simu}
In this section, We evaluate the numerical performance of the proposed CM--MRL procedure via simulations. We first consider a setting adapted from \citet{han2014multiply} to illustrate effectiveness and robustness under misspecification, and then compare against related conformal methods.

We generate i.i.d. covariates and corresponding outcomes
\begin{align*}
  &  X_1\sim \mathcal{N}(5,1),\quad
X_2\sim \operatorname{Bernoulli}(0.5),\quad
X_3\sim \mathcal{N}(0,1),\quad
X_4\sim \mathcal{N}(0,1), \\
&Y \;=\; 3.5 + 0.5 X_1 + 2 X_2 + X_3 + X_4 + \sigma\cdot\varepsilon_Y,
\end{align*} The true regression is linear with
$\beta_0=(3.5,0.5,2.0,1.0,1.0)$. The outcome follows a linear signal-plus-noise model
\[
Y \;=\; 3.5 + 0.5X_1 + 2X_2 + X_3 + X_4 + \sigma\,\varepsilon_Y,
\]
so that the conditional mean is $\mu_0(x)=\beta_0^\top x$ with $\beta_0=(3.5,\,0.5,\,2,\,1,\,1)$. Missingness follows the MAR propensity $
\operatorname{logit}\{\pi_0(X)\} \;=\; 3.5 - 5.0\,X_2,$
yielding approximately $58\%$ observed outcomes on average and about $42\%$ missing. We observe $R=\mathbbm{1}\{Y\ \text{observed}\}$ and $(X,R,YR)$. We vary the error law through three scenarios:
\begin{enumerate}
    \item[(A)] homoskedastic Gaussian noise, $\varepsilon_Y\sim\mathcal{N}(0,1)$ with $\sigma=1$;
    \item[(B)] heavy tails, $\varepsilon_Y\sim t_3$ scaled to unit variance with $\sigma=1$;
    \item[(C)] heteroskedasticity, $\varepsilon_Y\sim\mathcal{N}\!\big(0,\,(0.6+0.2|X_1|)^2\big)$ with $\sigma=1$ (variance rises with $|X_1|$).
\end{enumerate}
This variation of $(\sigma,\varepsilon_Y)$ vary by scenario is shown in Table \ref{choices}. 

To study model misspecification and robustness, we provide two candidate propensity models and two candidate outcome models:
\begin{align*}
\text{Propensity:}\quad
&\operatorname{logit}\{\pi^1(\alpha^1)\}=\alpha_1^1+\alpha_2^1 X_2\quad\text{(correct)},\\
&\operatorname{logit}\{\pi^2(\alpha^2)\}=\alpha_1^2+\alpha_2^2X_1+\alpha_3^2X_2+\alpha_4^2X_3+\alpha_5^2X_4\quad\text{(misspecified)};\\
\E(Y|X)\quad
&a^1(\gamma^1)=\gamma_1^1+\gamma_2^1 X_1+\gamma_3^1 X_2+\gamma_4^1 X_3+\gamma_5^1 X_4\quad\text{(correct)},\\
&a^2(\gamma^2)=\gamma_1^2+\gamma_2^2 X_1+\gamma_3^2 X_2+\gamma_4^2 X_3\quad\text{(misspecified)}.
\end{align*}
We examine four settings that mix which candidates are available to the learner shown in Table \ref{setting}.

\begin{figure}[t!]
\includegraphics [width=1\linewidth,height=4cm]{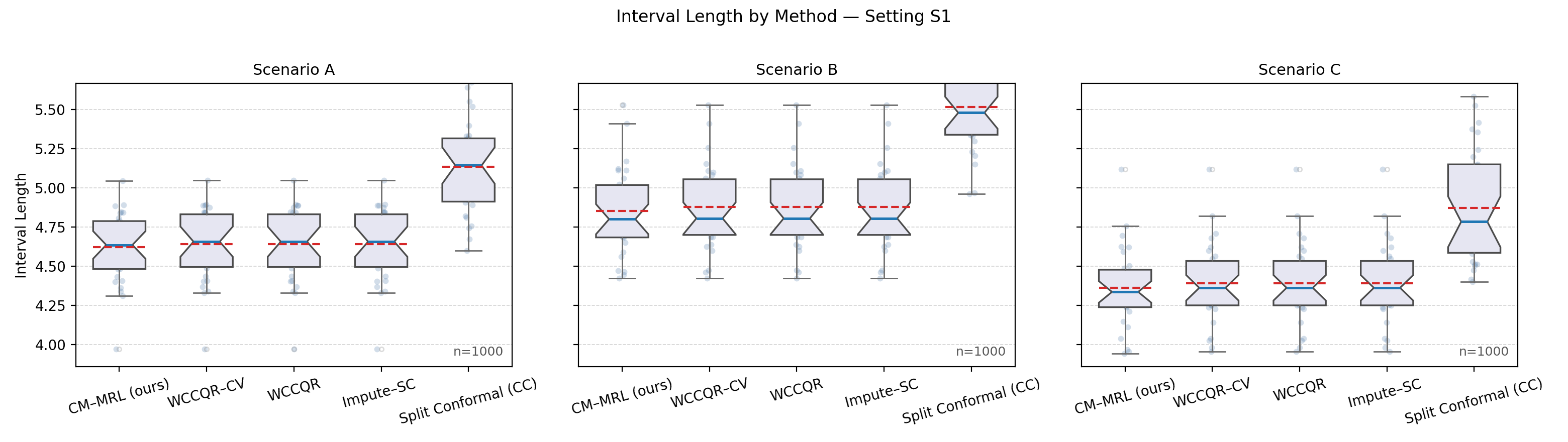}\par
\includegraphics[width=1\linewidth, height=4cm]{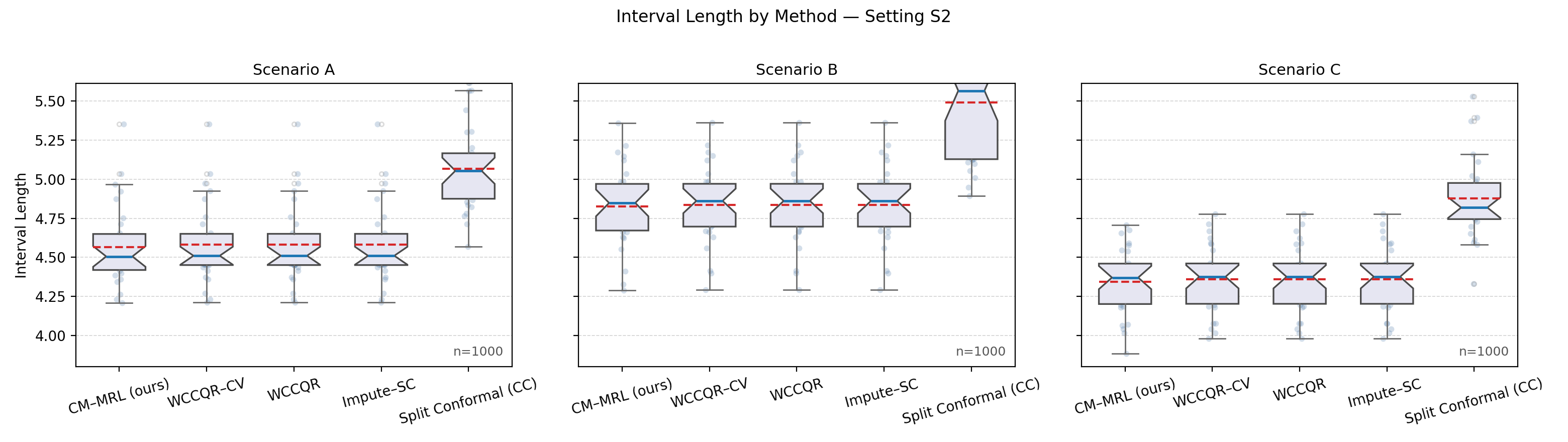} \par
\includegraphics[width=1\linewidth, height=4cm]{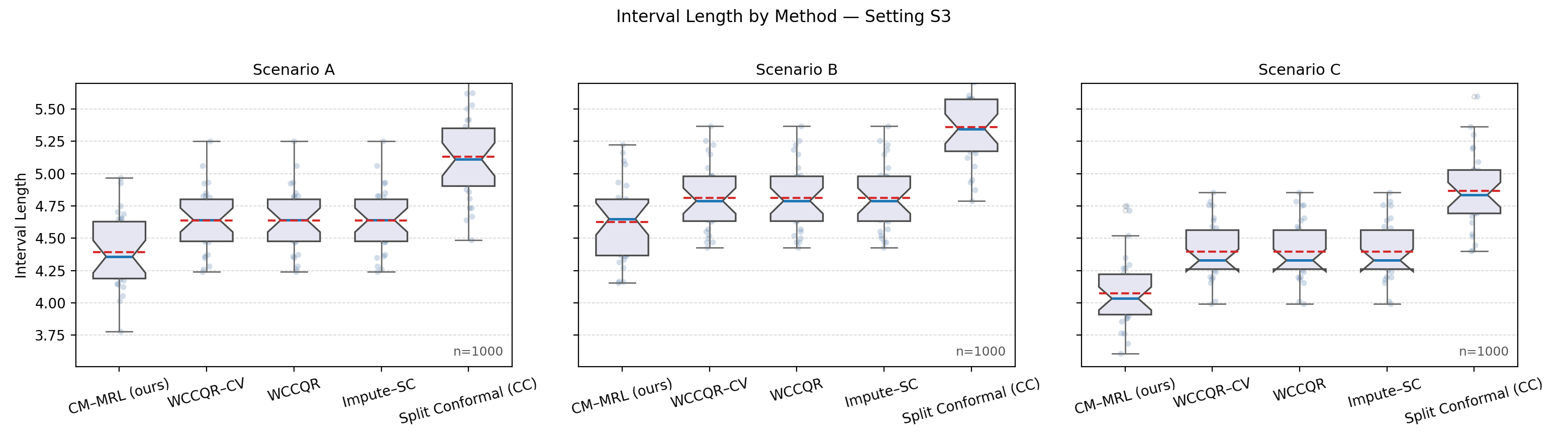} \par
\includegraphics[width=1\linewidth, height=4cm]{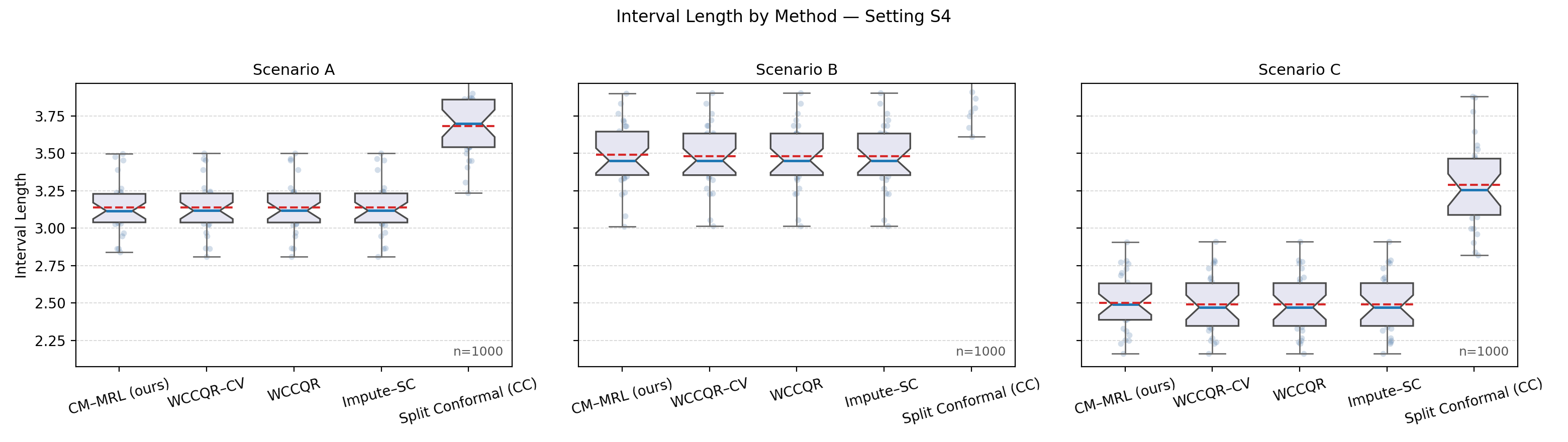} \par
%\centerline{\epsfig{file=d:/sinicas/simuh3.eps,angle=-90,width=4.5in}}\par
%\centerline{\epsfig{file=d:/sinicas/simuh4.eps,width=4.5in}}\par
	%\centerline{\epsfig{file=logo.eps,width=4.5in}}\par  %<- modified by Ivan  (You need to put figure in the same folder; or you must specify the path to the figure.)
\caption{Interval length of all methods under different settings and cadidate model selections.}
	\label{fig:result}
	\end{figure}

For each replicate we split the sample in half into training and calibration sets, fit candidate models on training, and construct prediction intervals on calibration at target marginal coverage $\tau=0.9$.
We compare our \textbf{CM--MRL} (double-calibrated, multiple-robust conformal) against: \textbf{Impute--SC} (impute then split conformal using the available $a^{(\cdot)}$),
complete-case \textbf{SC} (split conformal on observed pairs only), and the weighting baselines \textbf{WCCQR} and \textbf{WCCQR--CV}.
Performance is summarized over 50 Monte Carlo replications by empirical coverage and average interval length. The results shown in Table \ref{tab:summary_methods_stacked} reports mean (sd) and  Figure \ref{fig:result} reports boxplot of the length across all methods under different settings and scenarios.

Across all settings, complete–case split conformal (SC) systematically overcovers the nominal level $\tau=0.9$ and produces the longest intervals. In contrast, the three missing–data methods—\textbf{CM--MRL} (ours), Impute--SC, and WCCQR/WCCQR--CV—cluster near nominal coverage with markedly shorter lengths.
In settings S1 and S2 with correct $a^{(1)}$ available,
CM--MRL, Impute--SC, and WCCQR/WCCQR--CV achieve comparable coverage around $0.89$–$0.90$ with similar lengths, while SC is longer and overconservative. For example, in S1A, CM--MRL length is $4.621$ versus $5.133$ for SC; in S2C, CM--MRL is $4.344$ versus $4.878$ for SC. Standard deviations are small and similar across the missing–data methods.
In Setting S3 without no correct outcome model $a^{(1)}$, 
when the outcome working model is misspecified, CM--MRL delivers the {shortest} intervals among all methods, reflecting its calibration efficiency. SC again over-covers and is longest. This highlights a trade–off under misspecification: CM--MRL prioritizes shorter length while maintaining stable performance, whereas imputation/weighting baselines tilt toward slightly higher coverage with wider intervals.

In these simulations, WCCQR and WCCQR--CV track Impute--SC nearly identically in both coverage and length. This suggests that, under the MAR mechanism and the specific candidate models considered here, reweighting alone does not yield an advantage over the impute–then–conformal baseline, whereas the double calibration in CM--MRL can yield meaningful length reductions (S3/S4) with coverage close to nominal when at least one working model is well specified (S1/S2).
Therefore CM--MRL consistently produces intervals that are \emph{(i)} much shorter than complete–case SC across scenarios, and \emph{(ii)} competitive with or shorter than Impute--SC/WCCQR when a correct working model is present (S1/S2). Under outcome–model misspecification (S3), CM--MRL preserves its short–length advantage but exhibits mild undercoverage, illustrating the expected robustness/efficiency trade–off; in low–noise regimes (S4), it offers the best length–coverage balance among missing–data methods.

%The results are summarized based on 100 replications. The option of the models in each setting is specified in Table \ref{settings}.

\begin{table}[t!]
\centering
\caption{Choices of the error term $\varepsilon_Y$ and the scale $\sigma$.}
\label{choices}
{%
{\small\begin{tabular}{|c|l l|}
\hline
\textbf{No.} & $\sigma$ & $\varepsilon_Y$ \\ \hline
A & 1 & $\varepsilon_Y \sim \mathcal{N}(0,1)$ \\ \hline
B & 1 & $\varepsilon_Y \sim t_{3}$ \\ \hline
C & 0.6 & $\varepsilon_Y \sim \mathcal{N}\big(0,\,(0.6+0.2|X_1|)^2\big)$ \\ \hline
\end{tabular}%
}}
\\[4pt]
\parbox{0.95\linewidth}{\footnotesize
\emph{Note.} We use all three choices in every setting. $t_\nu$ denotes a Student-\emph{t} distribution with $\nu$ degrees of freedom. 
The probability of outlier occurrence is $0.02$ in choice 3.}
\end{table}

{\small\begin{table}[t!]
\centering
\caption{Three candidate propensity and imputation model combinations.}\label{setting}
{%
\begin{tabular}{|c|c c c c|}
\hline
\textbf{Setting} & {\bf$\pi^1\left(\boldsymbol{\alpha}^1\right)$} & {\bf $\pi^2\left(\boldsymbol{\alpha}^2\right)$} & {\bf $a^1\left(\boldsymbol{\gamma}^1\right)$} & {\bf $a^2\left(\boldsymbol{\gamma}^2\right)$} \\ \hline
 S1 & \cmark & \cmark & \cmark & \cmark \\\hline
			 S2 & \cmark & \xmark & \cmark & \cmark \\ \hline
S3& \cmark & \cmark & \xmark & \cmark \\\hline
S4& \cmark & \cmark & \cmark & \xmark \\\hline
\end{tabular}%
}
\\[4pt]
\parbox{0.95\linewidth}{\footnotesize
\emph{Note.} \cmark\quad indicates the model is adopted, while \xmark\quad indicates not adopted.}
\end{table}}

\begin{table}[t!]
\centering
\caption{Coverage and interval length (mean (sd)) across methods.}
\label{tab:summary_methods_stacked}
% compact formatting for this table only
\setlength{\tabcolsep}{4pt}\renewcommand{\arraystretch}{1.05}\scriptsize
\begin{tabular}{l *{6}{c}} % 1 method col + (A,B,C)×(Cov,Len) = 7 total
\toprule
\multicolumn{7}{c}{\textbf{Setting S1}}\\
\cmidrule(lr){1-7}
{Method}
  & \multicolumn{2}{c}{A} & \multicolumn{2}{c}{B} & \multicolumn{2}{c}{C} \\
\cmidrule(lr){2-3}\cmidrule(lr){4-5}\cmidrule(lr){6-7}
& \textbf{Cov.} & \textbf{Len.} & \textbf{Cov.} & \textbf{Len.} & \textbf{Cov.} & \textbf{Len.} \\
\midrule
CM--MRL
& 0.893(0.020) & 4.621(0.226) & 0.885(0.021) & 4.852(0.273) & 0.899(0.019) & 4.363(0.254) \\
Impute--SC
& 0.895(0.021) & 4.642(0.232) & 0.888(0.020) & 4.877(0.266) & 0.901(0.019) & 4.391(0.254) \\
Split Conformal (CC)
& 0.924(0.017) & 5.133(0.284) & 0.925(0.016) & 5.514(0.280) & 0.928(0.022) & 4.870(0.354) \\
WCCQR
& 0.895(0.021) & 4.642(0.232) & 0.888(0.020) & 4.877(0.266) & 0.901(0.019) & 4.391(0.254) \\
WCCQR--CV
& 0.895(0.021) & 4.642(0.232) & 0.888(0.020) & 4.877(0.266) & 0.901(0.019) & 4.391(0.254) \\
\midrule
\multicolumn{7}{c}{\textbf{Setting S2}}\\
\midrule
CM--MRL
& 0.887(0.021) & 4.565(0.256) & 0.886(0.021) & 4.826(0.253) & 0.896(0.020) & 4.344(0.209) \\
Impute--SC
& 0.889(0.021) & 4.580(0.251) & 0.887(0.021) & 4.836(0.245) & 0.897(0.020) & 4.360(0.211) \\
Split Conformal (CC)
& 0.923(0.018) & 5.066(0.257) & 0.927(0.021) & 5.491(0.355) & 0.927(0.016) & 4.878(0.253) \\
WCCQR
& 0.889(0.021) & 4.580(0.251) & 0.887(0.021) & 4.836(0.245) & 0.897(0.020) & 4.360(0.211) \\
WCCQR--CV
& 0.889(0.021) & 4.580(0.251) & 0.887(0.021) & 4.836(0.245) & 0.897(0.020) & 4.360(0.211) \\
\midrule
\multicolumn{7}{c}{\textbf{Setting S3}}\\
\midrule
CM--MRL
& 0.871(0.026) & 4.391(0.276) & 0.871(0.025) & 4.625(0.305) & 0.875(0.024) & 4.075(0.270) \\
Impute--SC
& 0.894(0.019) & 4.639(0.241) & 0.886(0.019) & 4.813(0.259) & 0.899(0.019) & 4.397(0.224) \\
Split Conformal (CC)
& 0.926(0.019) & 5.131(0.320) & 0.919(0.018) & 5.360(0.306) & 0.929(0.016) & 4.865(0.288) \\
WCCQR
& 0.894(0.019) & 4.639(0.241) & 0.886(0.019) & 4.813(0.259) & 0.899(0.019) & 4.397(0.224) \\
WCCQR--CV
& 0.894(0.019) & 4.639(0.241) & 0.886(0.019) & 4.813(0.259) & 0.899(0.019) & 4.397(0.224) \\
\midrule
\multicolumn{7}{c}{\textbf{Setting S4}}\\
\midrule
CM--MRL
& 0.879(0.025) & 3.139(0.169) & 0.872(0.025) & 3.490(0.214) & 0.877(0.019) & 2.501(0.189) \\
Impute--SC
& 0.879(0.025) & 3.139(0.171) & 0.871(0.025) & 3.480(0.214) & 0.875(0.019) & 2.491(0.191) \\
Split Conformal (CC)
& 0.930(0.017) & 3.683(0.213) & 0.923(0.018) & 4.187(0.282) & 0.933(0.015) & 3.291(0.285) \\
WCCQR
& 0.879(0.025) & 3.139(0.171) & 0.871(0.025) & 3.480(0.214) & 0.875(0.019) & 2.491(0.191) \\
WCCQR--CV
& 0.879(0.025) & 3.139(0.171) & 0.871(0.025) & 3.480(0.214) & 0.875(0.019) & 2.491(0.191) \\
\bottomrule
\end{tabular}
\end{table}

\subsection{Real Data}

As an application of the proposed method, we consider data from 2139 HIV-infected subjects enrolled in the AIDS Clinical Trials Group Protocol 175 (ACTG 175) \citep{hammer1996trial}. This randomized trial involved patients from 43 AIDS Clinical Trials Units and 9 National Hemophilia Foundation sites across the United States and Puerto Rico. The study assessed the efficacy of single versus dual nucleoside treatments in HIV-infected subjects with CD4 cell counts (a measure of immunologic status) between 200 and 500 per cubic millimeter. Subjects were randomly assigned to one of four antiretroviral regimens with equal probability: zidovudine (ZDV) alone, ZDV+didanosine (ddI), ZDV+zalcitabine (ddC), or ddI alone.
 
Following the analyses by \cite{han2014multiply}, we compare two treatment arms: the standard ZDV monotherapy arm and the combined arm of the three newer treatments. These two arms include 532 and 1607 subjects, respectively. Our primary focus is the effect of the treatment arm on CD4 counts measured at $96 \pm 5$ weeks post-baseline ($\mathrm{CD}4_{96}$), adjusting for baseline CD4 counts ($\mathrm{CD}4_0$) and other baseline characteristics. These characteristics include continuous covariates (age in years and weight in kilograms) and binary covariates (treatment, where $0=$ ZDV; race, where $0=$ white; gender, where $0=$ female; antiretroviral history, where $0=$ naive and $1=$ experienced; and whether the subject was off-treatment prior to 96 weeks, where $0=$ no). Our aim is to fit the following linear regression model:

$$
\begin{aligned}
\mathrm{CD}_{96} = & \beta_1 + \beta_2 \text{ trt } + \beta_3 \mathrm{CD}_0 + \beta_4 \text{ age } + \beta_5 \text{ weight } + \beta_6 \text{ race } \\
& + \beta_7 \text{ gender } + \beta_8 \text{ history } + \beta_9 \text{ offtrt } + \epsilon,
\end{aligned}
$$

where $\epsilon$ has a mean of zero conditional on all covariates. The data can be accessed through the R package \textit{speff2trial}\footnote{http://cran.r-project.org/web/packages/speff2trial/speff2trial.pdf}. The average age of the subjects is 35 years with a standard deviation of 8.7 years. The cohort includes 1522 white subjects and 617 nonwhite subjects. The subjects include 1171 males and 368 females. Among the patients, 1253 have a history of antiretroviral treatment, and 776 are off-treatment before 96 weeks. Due to some subjects dropping out of the study, CD4 counts at $96 \pm 5$ weeks are missing for 797 subjects (missingness rate of 37\%). 

Accurately specifying a model for $\pi(\boldsymbol{X})$ is challenging with an eight-dimensional $\boldsymbol{X}$, even with model selection and diagnostic techniques. The same challenge applies to modeling $\E(Y \mid \boldsymbol{X})$. Due to potential model misspecifications, the reliability of estimation and inference based on doubly robust methods can be questionable. Therefore, we apply the proposed method, which, although not definitive, may provide more reliable conclusions in the presence of model misspecifications, as demonstrated by the simulation studies in Section \ref{simu}.
Following \cite{han2014multiply}, we use a logistic regression model for 
propensity score function $\pi(\boldsymbol{X})$, and a linear regression model for $\E(Y \mid \boldsymbol{X})$. To ensure thoroughness, both models include all main effects of $\boldsymbol{X}$. We employ the estimating function $\boldsymbol{U}(Y, \boldsymbol{X}, \boldsymbol{\beta}) = \boldsymbol{X} \left(Y - \boldsymbol{X}^{\mathrm{T}} \boldsymbol{\beta}\right)$. Ultimately, our method achieved a coverage rate of 92.9\%, surpassing the target of 90\%. The resulting interval length was 196.413.

    %%%%%%%%%%%%%%%%%%%%%%%%%%%%%%%%%%%%%%%%%%%%%%%%%%%%%%%%%%%%%%%%%%%%%%%%%%%%%%%%%%%%%%%%%%%%%%%%%%%%%%%%%%%%%%%%%%%%%%%%%%%%

	%%%%%%%%%%%%%%%%%%%%%%%%%%%%%%%%%%%%%%%%%%%%%%%%%%%%%%%%%%%%%%%%%%%%%%%%%%%%%%%%%%%%%%%%%%

%\iffalse

\bibliographystyle{chicago}      % Chicago style, author-year citations
\bibliography{ref.bib}

\begin{thebibliography}{}

\bibitem[\protect\citeauthoryear{Cand{\`e}s, Lei, and Ren}{Cand{\`e}s
  et~al.}{2023}]{candes2023conformalized}
Cand{\`e}s, E., L.~Lei, and Z.~Ren (2023).
\newblock Conformalized survival analysis.
\newblock {\em Journal of the Royal Statistical Society Series B: Statistical
  Methodology\/}~{\em 85\/}(1), 24--45.

\bibitem[\protect\citeauthoryear{Cao, Tsiatis, and Davidian}{Cao
  et~al.}{2009}]{cao2009improving}
Cao, W., A.~A. Tsiatis, and M.~Davidian (2009).
\newblock Improving efficiency and robustness of the doubly robust estimator
  for a population mean with incomplete data.
\newblock {\em Biometrika\/}~{\em 96\/}(3), 723--734.

\bibitem[\protect\citeauthoryear{Feldman, Bates, and Romano}{Feldman
  et~al.}{2021}]{feldman2021improving}
Feldman, S., S.~Bates, and Y.~Romano (2021).
\newblock Improving conditional coverage via orthogonal quantile regression.
\newblock {\em Advances in neural information processing systems\/}~{\em 34},
  2060--2071.

\bibitem[\protect\citeauthoryear{Foygel~Barber, Candes, Ramdas, and
  Tibshirani}{Foygel~Barber et~al.}{2021}]{foygel2021limits}
Foygel~Barber, R., E.~J. Candes, A.~Ramdas, and R.~J. Tibshirani (2021).
\newblock The limits of distribution-free conditional predictive inference.
\newblock {\em Information and Inference: A Journal of the IMA\/}~{\em
  10\/}(2), 455--482.

\bibitem[\protect\citeauthoryear{Gibbs and Candes}{Gibbs and
  Candes}{2021}]{gibbs2021adaptive}
Gibbs, I. and E.~Candes (2021).
\newblock Adaptive conformal inference under distribution shift.
\newblock {\em Advances in Neural Information Processing Systems\/}~{\em 34},
  1660--1672.

\bibitem[\protect\citeauthoryear{Hammer, Katzenstein, Hughes, Gundacker,
  Schooley, Haubrich, Henry, Lederman, Phair, Niu, et~al.}{Hammer
  et~al.}{1996}]{hammer1996trial}
Hammer, S.~M., D.~A. Katzenstein, M.~D. Hughes, H.~Gundacker, R.~T. Schooley,
  R.~H. Haubrich, W.~K. Henry, M.~M. Lederman, J.~P. Phair, M.~Niu, et~al.
  (1996).
\newblock A trial comparing nucleoside monotherapy with combination therapy in
  hiv-infected adults with cd4 cell counts from 200 to 500 per cubic
  millimeter.
\newblock {\em New England Journal of Medicine\/}~{\em 335\/}(15), 1081--1090.

\bibitem[\protect\citeauthoryear{Han}{Han}{2014a}]{han2014further}
Han, P. (2014a).
\newblock A further study of the multiply robust estimator in missing data
  analysis.
\newblock {\em Journal of Statistical Planning and Inference\/}~{\em 148},
  101--110.

\bibitem[\protect\citeauthoryear{Han}{Han}{2014b}]{han2014multiply}
Han, P. (2014b).
\newblock Multiply robust estimation in regression analysis with missing data.
\newblock {\em Journal of the American Statistical Association\/}~{\em
  109\/}(507), 1159--1173.

\bibitem[\protect\citeauthoryear{Han}{Han}{2016}]{han2016intrinsic}
Han, P. (2016).
\newblock Intrinsic efficiency and multiple robustness in longitudinal studies
  with drop-out.
\newblock {\em Biometrika\/}~{\em 103\/}(3), 683--700.

\bibitem[\protect\citeauthoryear{Han, Kong, Zhao, and Zhou}{Han
  et~al.}{2019}]{han2019general}
Han, P., L.~Kong, J.~Zhao, and X.~Zhou (2019).
\newblock A general framework for quantile estimation with incomplete data.
\newblock {\em Journal of the Royal Statistical Society Series B: Statistical
  Methodology\/}~{\em 81\/}(2), 305--333.

\bibitem[\protect\citeauthoryear{Han and Wang}{Han and
  Wang}{2013}]{han2013estimation}
Han, P. and L.~Wang (2013).
\newblock Estimation with missing data: beyond double robustness.
\newblock {\em Biometrika\/}~{\em 100\/}(2), 417--430.

\bibitem[\protect\citeauthoryear{Kang and Schafer}{Kang and
  Schafer}{2007}]{kang2007demystifying}
Kang, J.~D. and J.~L. Schafer (2007).
\newblock Demystifying double robustness: A comparison of alternative
  strategies for estimating a population mean from incomplete data.

\bibitem[\protect\citeauthoryear{Lei, G’Sell, Rinaldo, Tibshirani, and
  Wasserman}{Lei et~al.}{2018}]{lei2018distribution}
Lei, J., M.~G’Sell, A.~Rinaldo, R.~J. Tibshirani, and L.~Wasserman (2018).
\newblock Distribution-free predictive inference for regression.
\newblock {\em Journal of the American Statistical Association\/}~{\em
  113\/}(523), 1094--1111.

\bibitem[\protect\citeauthoryear{Lei, Robins, and Wasserman}{Lei
  et~al.}{2013}]{lei2013distribution}
Lei, J., J.~Robins, and L.~Wasserman (2013).
\newblock Distribution-free prediction sets.
\newblock {\em Journal of the American Statistical Association\/}~{\em
  108\/}(501), 278--287.

\bibitem[\protect\citeauthoryear{Lei and Wasserman}{Lei and
  Wasserman}{2014}]{lei2014distribution}
Lei, J. and L.~Wasserman (2014).
\newblock Distribution-free prediction bands for non-parametric regression.
\newblock {\em Journal of the Royal Statistical Society: Series B: Statistical
  Methodology\/}, 71--96.

\bibitem[\protect\citeauthoryear{Lei and Cand{\`e}s}{Lei and
  Cand{\`e}s}{2021}]{lei2021conformal}
Lei, L. and E.~J. Cand{\`e}s (2021).
\newblock Conformal inference of counterfactuals and individual treatment
  effects.
\newblock {\em Journal of the Royal Statistical Society Series B: Statistical
  Methodology\/}~{\em 83\/}(5), 911--938.

\bibitem[\protect\citeauthoryear{Li, Gu, and Liu}{Li
  et~al.}{2020}]{li2020demystifying}
Li, W., Y.~Gu, and L.~Liu (2020).
\newblock Demystifying a class of multiply robust estimators.
\newblock {\em Biometrika\/}~{\em 107\/}(4), 919--933.

\bibitem[\protect\citeauthoryear{Little and Rubin}{Little and
  Rubin}{2002}]{little2002maximum}
Little, R.~J. and D.~B. Rubin (2002).
\newblock Maximum likelihood for general patterns of missing data: Introduction
  and theory with ignorable nonresponse.
\newblock {\em Statistical analysis with missing data\/}, 164--189.

\bibitem[\protect\citeauthoryear{Little and Rubin}{Little and
  Rubin}{2019}]{little2019statistical}
Little, R.~J. and D.~B. Rubin (2019).
\newblock {\em Statistical analysis with missing data}, Volume 793.
\newblock John Wiley \& Sons.

\bibitem[\protect\citeauthoryear{Robins and Rotnitzky}{Robins and
  Rotnitzky}{1995}]{robins1995semiparametric}
Robins, J.~M. and A.~Rotnitzky (1995).
\newblock Semiparametric efficiency in multivariate regression models with
  missing data.
\newblock {\em Journal of the American Statistical Association\/}~{\em
  90\/}(429), 122--129.

\bibitem[\protect\citeauthoryear{Romano, Patterson, and Candes}{Romano
  et~al.}{2019}]{romano2019conformalized}
Romano, Y., E.~Patterson, and E.~Candes (2019).
\newblock Conformalized quantile regression.
\newblock {\em Advances in Neural Information Processing Systems\/}~{\em 32},
  3543--3553.

\bibitem[\protect\citeauthoryear{Rosenbaum and Rubin}{Rosenbaum and
  Rubin}{1983}]{rosenbaum1983central}
Rosenbaum, P.~R. and D.~B. Rubin (1983).
\newblock The central role of the propensity score in observational studies for
  causal effects.
\newblock {\em Biometrika\/}~{\em 70\/}(1), 41--55.

\bibitem[\protect\citeauthoryear{Rubin}{Rubin}{1996}]{rubin1996multiple}
Rubin, D.~B. (1996).
\newblock Multiple imputation after 18+ years.
\newblock {\em Journal of the American statistical Association\/}~{\em
  91\/}(434), 473--489.

\bibitem[\protect\citeauthoryear{Rubin}{Rubin}{2018}]{rubin2018multiple}
Rubin, D.~B. (2018).
\newblock Multiple imputation.
\newblock In {\em Flexible Imputation of Missing Data, Second Edition}, pp.\
  29--62. Chapman and Hall/CRC.

\bibitem[\protect\citeauthoryear{Sesia and Romano}{Sesia and
  Romano}{2021}]{sesia2021conformal}
Sesia, M. and Y.~Romano (2021).
\newblock Conformal prediction using conditional histograms.
\newblock {\em Advances in Neural Information Processing Systems\/}~{\em 34},
  6304--6315.

\bibitem[\protect\citeauthoryear{Tan}{Tan}{2010}]{tan2010bounded}
Tan, Z. (2010).
\newblock Bounded, efficient and doubly robust estimation with inverse
  weighting.
\newblock {\em Biometrika\/}~{\em 97\/}(3), 661--682.

\bibitem[\protect\citeauthoryear{Tibshirani, Foygel~Barber, Candes, and
  Ramdas}{Tibshirani et~al.}{2019}]{tibshirani2019conformal}
Tibshirani, R.~J., R.~Foygel~Barber, E.~Candes, and A.~Ramdas (2019).
\newblock Conformal prediction under covariate shift.
\newblock {\em Advances in neural information processing systems\/}~{\em 32}.

\bibitem[\protect\citeauthoryear{Tu, Zhang, Ackermann, Mohan, Kjellstr{\"o}m,
  and Zhang}{Tu et~al.}{2019}]{tu2019causal}
Tu, R., C.~Zhang, P.~Ackermann, K.~Mohan, H.~Kjellstr{\"o}m, and K.~Zhang
  (2019).
\newblock Causal discovery in the presence of missing data.
\newblock In {\em The 22nd International Conference on Artificial Intelligence
  and Statistics}, pp.\  1762--1770. PMLR.

\bibitem[\protect\citeauthoryear{Vovk}{Vovk}{2012}]{vovk2012conditional}
Vovk, V. (2012).
\newblock Conditional validity of inductive conformal predictors.
\newblock In {\em Asian conference on machine learning}, pp.\  475--490. PMLR.

\bibitem[\protect\citeauthoryear{Vovk, Gammerman, and Shafer}{Vovk
  et~al.}{2005}]{vovk2005algorithmic}
Vovk, V., A.~Gammerman, and G.~Shafer (2005).
\newblock {\em Algorithmic learning in a random world}.
\newblock Springer Science \& Business Media.

\bibitem[\protect\citeauthoryear{Vovk, Petej, Toccaceli, Gammerman, Ahlberg,
  and Carlsson}{Vovk et~al.}{2020}]{vovk2020conformal}
Vovk, V., I.~Petej, P.~Toccaceli, A.~Gammerman, E.~Ahlberg, and L.~Carlsson
  (2020).
\newblock Conformal calibrators.
\newblock In {\em conformal and probabilistic prediction and applications},
  pp.\  84--99. PMLR.

\bibitem[\protect\citeauthoryear{Zaffran, Dieuleveut, Josse, and
  Romano}{Zaffran et~al.}{2023}]{zaffran2023conformal}
Zaffran, M., A.~Dieuleveut, J.~Josse, and Y.~Romano (2023).
\newblock Conformal prediction with missing values.
\newblock {\em arXiv preprint arXiv:2306.02732\/}.

\bibitem[\protect\citeauthoryear{Zaffran, F{\'e}ron, Goude, Josse, and
  Dieuleveut}{Zaffran et~al.}{2022}]{zaffran2022adaptive}
Zaffran, M., O.~F{\'e}ron, Y.~Goude, J.~Josse, and A.~Dieuleveut (2022).
\newblock Adaptive conformal predictions for time series.
\newblock In {\em International Conference on Machine Learning}, pp.\
  25834--25866. PMLR.

\end{thebibliography}

\end{document}